\documentclass[a4paper,aps,prd,10pt,preprintnumbers,showpacs,twocolumn,superscriptaddress,nofootinbib,amsmath,amssymb]{revtex4-1}
\usepackage{graphicx}
\usepackage{cmap}
\usepackage[utf8]{inputenc}
\usepackage[T1]{fontenc}

\begin{document}
\title{Long-Lived Quasinormal Modes of Brane-Localized Reissner--Nordström--de Sitter Black Holes}
\author{Zainab Malik}\email{zainabmalik8115@outlook.com}
\affiliation{Institute of Applied Sciences and Intelligent Systems, H-15, Pakistan}
\begin{abstract}
We study the quasinormal modes of a massive scalar field propagating on the Reissner--Nordström--de Sitter (RNdS) black hole black hole background on a 3+1-dimensional brane embedded in a higher-dimensional world. Using the WKB method supplemented with Padé approximants and validated by time-domain integration via the Prony method, we compute the dominant quasinormal frequencies for a wide range of black hole and field parameters. We show that the presence of the cosmological constant, black hole charge, and bulk dimensionality significantly affect the oscillation frequencies and damping rates of the scalar perturbations. In particular, we observe the emergence of long lived modes and slowly decaying oscillatory tails in the regime of large field mass. The results demonstrate good agreement between the frequency- and time-domain methods, reinforcing the reliability of the semi-analytic approach in this context.
\end{abstract}

\maketitle
\section{Introduction}

Quasinormal modes (QNMs) represent the characteristic oscillations of black holes and other compact objects under external perturbations~\cite{Kokkotas:1999bd,Berti:2009kk,Konoplya:2011qq,Bolokhov:2025uxz}. These complex-frequency solutions, governed by boundary conditions of purely ingoing waves at the horizon and purely outgoing waves at infinity (or regularity at the cosmological horizon in de Sitter space), encode essential information about the stability and dynamical response of the underlying spacetime. In recent years, the study of QNMs has gained renewed relevance due to the detection of gravitational waves \cite{LIGOScientific:2016aoc, LIGOScientific:2017vwq, LIGOScientific:2020zkf, Babak:2017tow}, in which the ringdown signal can be modeled by quasinormal ringing, offering a direct test of general relativity in the strong-field regime.

The investigation of QNMs in higher-dimensional black hole backgrounds has attracted significant interest \cite{Seahra:2004fg,Seahra:2005wk,Kodama:2009bf,Witek:2010xi,Emparan:2015rva,Zhidenko:2006rs,Liu:2019lon,Cuyubamba:2016cug,Konoplya:2013sba,Destounis:2019hca,Mishra:2021waw} in the context of brane-world scenarios and string-inspired models of gravity~\cite{Maartens:2010ar,Emparan:2008eg}. In these models, our observable universe is described as a 3+1-dimensional brane embedded in a higher-dimensional bulk spacetime. While the full gravitational dynamics occur in the higher-dimensional manifold, standard model fields are typically confined to the brane. A natural and phenomenologically relevant question arises: how do scalar fields behave when propagating on the brane in the background of a higher-dimensional black hole, such as the Reissner–Nordström–de Sitter (RNdS) solution? In particular, the presence of charge and a cosmological constant introduces rich structure in the causal geometry, including multiple horizons, which significantly affect the quasinormal spectrum.

Quasinormal modes of massless Standard Model fields have been studied extensively for asymptotically flat black holes, including those with a nonzero cosmological constant~\cite{Kanti:2006ua,Kanti:2005xa}. However, to the best of our knowledge, no such analyses exist for massive fields in this context, except the case of vanishing charge and cosmological constant \cite{Zinhailo:2024jzt}. In this work, we aim to fill this gap by studying the quasinormal modes of a massive scalar field propagating on the brane, in black hole spacetimes with both a nonvanishing cosmological constant and electric charge.

Massive scalar fields play a particularly intriguing role in the study of black hole perturbations. Unlike their massless counterparts, massive fields exhibit distinctive features both in the frequency domain and in time-domain evolution. One of the key phenomena is the existence of \emph{quasi-resonances}—arbitrarily long-lived modes ~\cite{Ohashi:2004wr,Zhidenko:2006rs}. Quasi-resonances take place for various black hole and wormhole models and types of fields \cite{Konoplya:2017tvu,Bolokhov:2023bwm,Churilova:2019qph,Fernandes:2021qvr,Percival:2020skc,Zinhailo:2018ska,Lutfuoglu:2025hwh}, still they are not guaranteed and there are exceptions when the quasi-resonances do not exist \cite{Zinhailo:2024jzt,Lutfuoglu:2025hjy}

In the time domain, massive fields are characterized by late-time tails that are not purely decaying, but oscillatory with a power-law envelope \cite{Jing:2004zb,Koyama:2001qw,Moderski:2001tk,Rogatko:2007zz,Koyama:2001ee,Konoplya:2006gq}. These slowly decaying tails may persist well beyond the ringdown phase, potentially leading to observable signatures in the Pulsar Timing Array Experiment or future gravitational wave data \cite{NANOGrav:2020bcs, NANOGrav:2023hvm, Seahra:2004fg, Konoplya:2023fmh}.

In particular, such massive tails may be detectable through observations using Pulsar Timing Arrays (PTAs), which are sensitive to long-duration, low-frequency gravitational signals~\cite{NANOGrav:2020bcs}. Furthermore, even originally massless fields may acquire an \emph{effective mass} in certain physical scenarios. For instance, a scalar field propagating in a black hole spacetime immersed in a magnetic field can behave effectively as a massive field due to interactions with the background electromagnetic field~\cite{Konoplya:2008hj,Kokkotas:2010zd,Konoplya:2007yy,Davlataliev:2024mjl,Wu:2015fwa}. These effects not only enrich the phenomenology of black hole perturbations but also offer new avenues for probing the properties of higher-dimensional spacetimes and quantum gravity corrections.

Quasinormal modes of asymptotically de Sitter black holes have attracted significant attention due to both theoretical and observational motivations. The presence of a cosmological horizon introduces a second natural boundary, modifying the boundary conditions and leading to a discrete spectrum that governs the decay of perturbations at all time scales \cite{Dyatlov:2010hq,Dyatlov:2011jd,Dubinsky:2024jqi}. These modes play a crucial role in studies of the Strong Cosmic Censorship conjecture \cite{Cardoso:2017soq,Dias:2018ynt,Dias:2018ufh,Chrysostomou:2025qud,Chrysostomou:2025twu,Mo:2018nnu,Ge:2018vjq,Hod:2018lmi}, stability of black hole spacetimes with a positive cosmological constant \cite{Cuyubamba:2016cug,Konoplya:2013sba,Konoplya:2017lhs,Konoplya:2017ymp}, and potential signatures in early-universe or inflationary scenarios. Moreover, de Sitter backgrounds arise naturally in many models of dark energy and modified gravity, making the understanding of their quasinormal spectra both timely and relevant. As a result, a great number of publications are dedicated to quasinormal modes of asymptotically de Sitter black holes (see, for instance, \cite{Zhidenko:2003wq,Konoplya:2007zx,Jing:2003wq,Aragon:2020qdc,Mo:2018nnu,Konoplya:2004uk,Dubinsky:2024hmn}).

The paper is organized as follows. In Sec.~\ref{sec:background}, we review the higher-dimensional Reissner--Nordström--de Sitter (RNdS) black hole and its projection onto the 3+1-dimensional brane. In addition, there we present the effective potential for massive scalar perturbations and the master wave equation governing their evolution. In Sec.~\ref{sec:WKB} and \ref{sec:timedomain}, we describe the computational techniques used to extract the quasinormal frequencies, including the WKB method with Padé approximants and the time-domain integration using the Prony method. Sec.~\ref{sec:QNMs} is devoted to a detailed analysis of the quasinormal spectra for various dimensions, field masses, and black hole parameters. Finally, in Sec.~\ref{sec:conclusions}, we summarize our main findings and outline possible directions for future research.

\section{Black hole metric and effective potential.}\label{sec:background}

The Tangherlini solution \cite{Tangherlini:1963bw} generalizes the Schwarzschild or Reissner--Nordström black hole to higher dimensions. The full \( D \)-dimensional Reissner--Nordström--de Sitter (RNdS) spacetime is described by the line element
\begin{equation}
ds^2 = -f(r) dt^2 + \frac{dr^2}{f(r)} + r^2 d\Omega_{D-2}^2,
\label{eq:fullmetric}
\end{equation}
where \( d\Omega_{D-2}^2 \) denotes the line element on the unit \((D-2)\)-sphere, and the metric function \( f(r) \) is given by
\begin{equation}
f(r) = 1 - \frac{2 \Lambda r^2}{(D-1)(D-2)} - \frac{2M}{r^{D-3}} + \frac{Q^2}{r^{2D - 6}},
\label{eq:metricfunction}
\end{equation}
where $M$ is the black hole mass, $Q$ is the electric charge and $\Lambda$ is the cosmological constant. 

In brane-world scenarios, standard model fields are localized on a 3+1-dimensional hypersurface embedded in the higher-dimensional bulk. When the size of extra dimension is much larger than the black hole size, the black hole metric can be approximated by the  Tangherlini solution. To study field propagation on the brane, one projects the bulk geometry onto a 4-dimensional slice by restricting the angular part of the metric to a 2-sphere. This effectively replaces the full angular sector \( d\Omega_{D-2}^2 \) with the usual
\begin{equation}
d\Omega_{2}^2 = d\theta^2 + \sin^2\theta\, d\phi^2.
\end{equation}

As a result, the brane-projected line element takes the form
\begin{equation}
ds^2 = -f(r) dt^2 + \frac{dr^2}{f(r)} + r^2 \left( d\theta^2 + \sin^2\theta\, d\phi^2 \right),
\label{eq:projectedmetric}
\end{equation}
where the metric function \( f(r) \) retains its dependence on the higher-dimensional parameters \( M \), \( Q \), \( \Lambda \), and \( D \), as given in Eq.~\eqref{eq:metricfunction}. While this projection does not produce a fully consistent lower-dimensional solution to the Einstein equations, it captures the effective influence of the higher-dimensional geometry on brane-localized fields and has been widely used in the literature to study black hole perturbations in brane-world models \cite{Kanti:2005xa,Konoplya:2017ymp,Kanti:2006ua,Kanti:2004nr}.

Finally, the metric of the brane-localized RNdS black hole is given by the following line element,
\begin{equation}\label{metric}
  ds^2=-f(r)dt^2+\frac{dr^2}{f(r)}+r^2(d\theta^2+\sin^2\theta d\phi^2),
\end{equation}
where $f(r)$ is given by eq. \ref{eq:metricfunction}. We will  measure all dimensional quantities in the units of mass $M=1$.

The dynamics of a scalar field in a curved spacetime background are governed by the generally covariant Klein–Gordon equation, which for a field of mass $\mu$ takes the form:
\begin{equation}\label{KGg}
\frac{1}{\sqrt{-g}} \partial_\mu \left( \sqrt{-g} \, g^{\mu\nu} \partial_\nu \Phi \right) = \mu^2 \Phi,
\end{equation}
where $g$ is the determinant of the metric tensor $g_{\mu\nu}$ and $\Phi$ represents the scalar field. Assuming a spherically symmetric and static background metric of the form~(\ref{metric}), one can employ separation of variables to decompose the scalar field into temporal, radial, and angular parts. As a result, the radial component satisfies a one-dimensional Schrödinger-like wave equation of the form~\cite{Kokkotas:1999bd,Berti:2009kk,Konoplya:2011qq}:
\begin{equation}\label{wave-equation}
\frac{d^2 \Psi}{dr_*^2} + \left( \omega^2 - V(r) \right) \Psi = 0,
\end{equation}
where $\omega$ is the oscillation frequency and $r_*$ denotes the tortoise coordinate, defined in terms of the radial coordinate by:
\begin{equation}\label{tortoise}
\frac{dr_*}{dr} = \frac{1}{f(r)}.
\end{equation}

The form of the effective potential $V(r)$ depends on the nature of the perturbing field. For scalar ($s = 0$) and electromagnetic ($s = 1$) fields, the effective potential reads:
\begin{equation}\label{potentialScalar}
V(r) = f(r) \left[ \frac{\ell(\ell+1)}{r^2} + \mu^2 \right] + \frac{1}{r} \frac{d^2 r}{dr_*^2},
\end{equation}
where $\ell$ is the angular momentum quantum number (multipole number), taking integer values $\ell = 0, 1, 2, \ldots$.

It is important to note that in this asymptotically de Sitter spacetimes, the effective potential is positive-definite in the entire domain between the event horizon and the cosmological horizon. This feature ensures the linear stability of the spacetime under scalar and electromagnetic perturbations, as it prohibits the existence of bound states with negative energy and consequently excludes exponentially growing modes.

\begin{figure}
\resizebox{\linewidth}{!}{\includegraphics{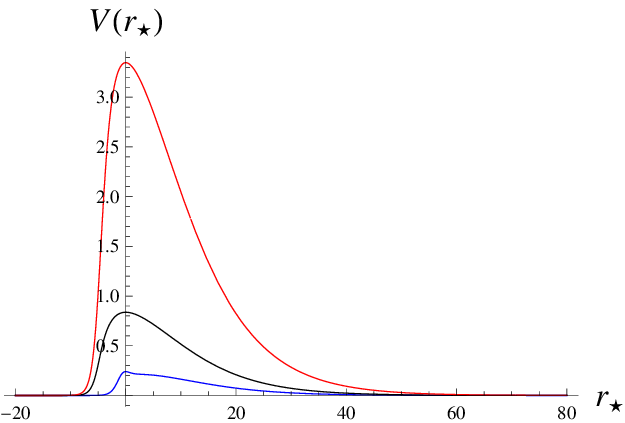}}
\caption{Effective potential as a function of the tortoise coordinate $r^{*}$ for $\ell=0$ scalar field perturbations: $\Lambda=0.02$, $Q=0.1$, $M=1$, $\mu=0.5$ (blue), $\mu=1$ (black) and $\mu=2$ (red).}\label{fig:scalarpot}
\end{figure}

\begin{figure}
\resizebox{\linewidth}{!}{\includegraphics{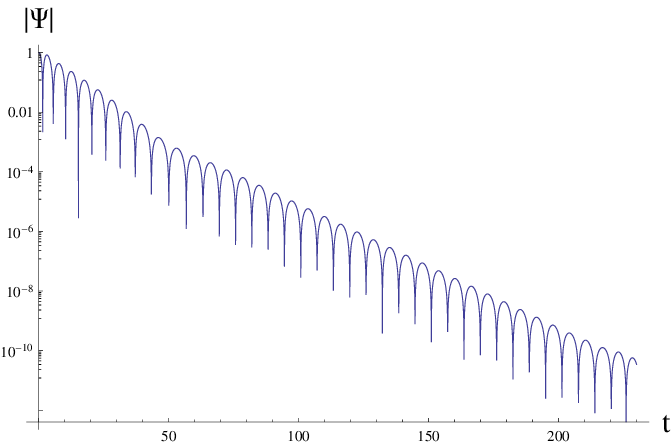}}
\caption{Time-domain profile for $\ell=1$ perturbations: $\Lambda=0.01$, $Q=0.01$, $M=1$, $\mu=0.5$. The Prony method via fitting at late times gives the fundamental mode $\omega=0.50021 - 0.09534 i$, while the fitting at the early ringdown period gives the mode of the Schwarzschild branch  $\omega = 0.604 - 0.189 i $, which is very close to the WKB value $\omega = 0.608902 - 0.188522 i$. 
}\label{fig:TD}
\end{figure}

\begin{figure}
\resizebox{\linewidth}{!}{\includegraphics{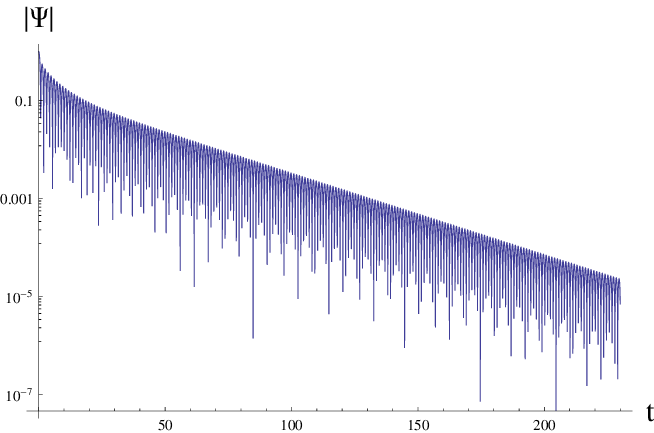}}
\caption{Time-domain profile for $\ell=1$ perturbations: $\Lambda=0.01$, $Q=0.01$, $M=1$, $\mu=3$. The Prony method via fitting at late times gives the fundamental mode  $\omega=2.832192- 0.0383320 i$, and the  first overtones $\omega = 2.8343531 -0.117965 i $, which is very close to the WKB values  $\omega = 2.831603 - 0.038356 i $ and $\omega = 2.831604-0.115066 i$.}\label{fig:TD2}
\end{figure}

\section{WKB Method and Padé Approximants}
\label{sec:WKB}

The Wentzel–Kramers–Brillouin (WKB) approximation is a widely used semi-analytic method for computing quasinormal modes of black holes~\cite{Schutz:1985km,Iyer:1986np,Konoplya:2003ii}. It is particularly effective for potentials that possess a single peak and vanish at the boundaries, such as those arising in asymptotically flat or de Sitter spacetimes. The WKB method is formulated by expanding the wave equation near the peak of the effective potential, treating the problem as a barrier penetration problem analogous to quantum tunneling.

In the $N$th-order WKB approximation, the quasinormal frequencies $\omega$ satisfy the condition
\begin{equation}
i \frac{Q_0}{\sqrt{2 Q_0''}} - \sum_{j=2}^{N} \Lambda_j = n + \frac{1}{2}, \quad n = 0,1,2,\dots,
\label{eq:WKB}
\end{equation}
where \( Q(r) = \omega^2 - V(r) \), \( Q_0 \) is the value of \( Q(r) \) at the peak of the effective potential, \( Q_0'' \) is the second derivative with respect to the tortoise coordinate \( r_* \) evaluated at the peak, and \( \Lambda_j \) are the higher-order correction terms determined recursively.

While increasing the WKB order generally improves the accuracy, the series does not always converge well—especially for low multipole numbers or near-extremal backgrounds. To overcome this, Padé approximants can be applied to the WKB series, effectively resumming it into a rational function form that better captures the underlying analytic structure.

Let the WKB expansion for \( \omega^2 \) be denoted by a truncated series:
\begin{equation}
\omega^2 = \sum_{k=0}^{N} a_k,
\end{equation}
where \( a_k \) are WKB terms depending on derivatives of the potential and \( n \). The Padé approximant of order \( P_m^n \) is defined as
\begin{equation}
P_m^n(x) = \frac{a_0 + a_1 x + \dots + a_m x^m}{1 + b_1 x + \dots + b_n x^n},
\end{equation}
constructed such that its Taylor expansion matches the WKB series up to order \( m+n \). In practice, the most commonly used Padé approximants for quasinormal mode calculations are \( P_6^6 \), \( P_7^6 \), or \( P_6^7 \), depending on the order of the WKB correction terms available.

This method, known as the \textit{Padé-WKB approach}, has been shown to significantly improve the accuracy and stability of quasinormal frequency calculations, even for low values of \( \ell \) or near-extremal black holes~\cite{Matyjasek:2017psv}. It provides excellent agreement with fully numerical methods such as Leaver’s continued fraction technique or time-domain integration, particularly when the effective potential has a single, smooth peak.

In this work, we employ the 6th-order WKB method supplemented with Padé approximants to compute the quasinormal modes of the massive scalar field on the brane in a higher-dimensional RNdS black hole background. This allows us to extract the dominant frequencies with high precision across a wide range of parameter values.

The explicit expressions for the higher-order WKB correction terms \( \Lambda_i \) are available in a series of works. The second- and third-order terms were originally derived in~\cite{Iyer:1986np}, while the fourth through sixth orders were developed in~\cite{Konoplya:2003ii}. More recently, the WKB expansion was extended up to the 13th order in~\cite{Matyjasek:2017psv}. Comprehensive discussions and practical applications of these corrections can also be found in review articles such as~\cite{Konoplya:2019hlu,Bolokhov:2025uxz}.

The WKB method, in both its standard and Padé-resummed forms, has become a standard tool for computing quasinormal modes and grey-body factors in a wide range of gravitational backgrounds. It has been applied across numerous contexts and spacetime geometries, including higher-dimensional black holes, non-minimally coupled fields, modified gravity theories, and black holes with nontrivial matter content~\cite{Konoplya:2005sy,Fu:2022cul,Yu:2022yyv,Konoplya:2001ji,Churilova:2021tgn,Konoplya:2022hll,Zinhailo:2019rwd,Guo:2020blq,Malik:2024tuf,Malik:2023bxc,Malik:2024sxv,DuttaRoy:2022ytr,Paul:2023eep,Dubinsky:2024fvi,Tan:2022vfe,Malik:2024elk}.
\begin{table}
\centering
{\scriptsize
\begin{tabular}{c c c c c}
\hline
Q & $\mu$ & WKB6 $m=3$ & WKB8 $m=4$ & difference  \\
\hline
$0.01$ & $0.1$ & $0.200425-0.279844 i$ & $0.204634-0.284449 i$ & $1.81\%$\\
$0.01$ & $0.5$ & $0.318088-0.123095 i$ & $0.172053-0.280303 i$ & $62.9\%$\\
$0.01$ & $1.$ & $0.942864-0.037956 i$ & $0.942779-0.037954 i$ & $0.00904\%$\\
$0.01$ & $1.5$ & $1.412235-0.038235 i$ & $1.412239-0.038227 i$ & $0.00062\%$\\
$0.01$ & $2.$ & $1.882104-0.038300 i$ & $1.882105-0.038300 i$ & $0.00003\%$\\
$0.01$ & $2.5$ & $2.352128-0.038335 i$ & $2.352128-0.038335 i$ & $0\%$\\
$0.01$ & $3.$ & $2.822226-0.038354 i$ & $2.822226-0.038354 i$ & $0\%$\\
$0.5$ & $0.1$ & $0.209554-0.275689 i$ & $0.214413-0.280224 i$ & $1.92\%$\\
$0.5$ & $0.5$ & $0.325734-0.130310 i$ & $0.187930-0.273660 i$ & $56.7\%$\\
$0.5$ & $1.$ & $0.942957-0.037905 i$ & $0.942864-0.037901 i$ & $0.00989\%$\\
$0.5$ & $1.5$ & $1.412382-0.038175 i$ & $1.412386-0.038167 i$ & $0.00069\%$\\
$0.5$ & $2.$ & $1.882311-0.038238 i$ & $1.882312-0.038237 i$ & $0.00004\%$\\
$0.5$ & $2.5$ & $2.352393-0.038271 i$ & $2.352393-0.038271 i$ & $3.4\times 10^{-6}\%$\\
$0.5$ & $3.$ & $2.822549-0.038289 i$ & $2.822549-0.038289 i$ & $0\%$\\
$0.99$ & $0.1$ & $0.229464-0.230300 i$ & $0.228462-0.232015 i$ & $0.611\%$\\
$0.99$ & $0.5$ & $0.351808-0.125562 i$ & $0.210178-0.226338 i$ & $46.5\%$\\
$0.99$ & $1.$ & $0.943239-0.037751 i$ & $0.943109-0.037742 i$ & $0.0138\%$\\
$0.99$ & $1.5$ & $1.412811-0.037997 i$ & $1.412817-0.037986 i$ & $0.00088\%$\\
$0.99$ & $2.$ & $1.882920-0.038048 i$ & $1.882921-0.038048 i$ & $0.00005\%$\\
$0.99$ & $2.5$ & $2.353176-0.038077 i$ & $2.353176-0.038077 i$ & $5.9\times 10^{-6}\%$\\
$0.99$ & $3.$ & $2.823503-0.038093 i$ & $2.823503-0.038093 i$ & $0\%$\\
\hline
\end{tabular}
}
\caption{Fundamental ($n=0$) quasinormal modes of the $\ell=0$ test scalar field for the RNdS black hole ($M=1$, $\Lambda=0.01$) calculated using the WKB formula at different orders and Padé approximants; $D=5$.}\label{table1}
\end{table}
\begin{table}
\centering
{\scriptsize
\begin{tabular}{c c c c c}
\hline
Q & $\mu$ & WKB6 $m=3$ & WKB8 $m=4$ & difference  \\
\hline
$0.01$ & $0.1$ & $0.529695-0.253667 i$ & $0.529738-0.253573 i$ & $0.0176\%$\\
$0.01$ & $0.5$ & $0.608902-0.188522 i$ & $0.609797-0.188666 i$ & $0.142\%$\\
$0.01$ & $1.$ & $0.982740-0.048021 i$ & $0.982740-0.048021 i$ & $0\%$\\
$0.01$ & $1.5$ & $1.433674-0.038174 i$ & $1.433718-0.038160 i$ & $0.00326\%$\\
$0.01$ & $2.$ & $1.896893-0.038311 i$ & $1.896898-0.038314 i$ & $0.00029\%$\\
$0.01$ & $2.5$ & $2.363572-0.038339 i$ & $2.363573-0.038339 i$ & $9.2\times 10^{-6}\%$\\
$0.01$ & $3.$ & $2.831603-0.038356 i$ & $2.831603-0.038356 i$ & $0\%$\\
$0.5$ & $0.1$ & $0.541429-0.251495 i$ & $0.541462-0.251134 i$ & $0.0607\%$\\
$0.5$ & $0.5$ & $0.618030-0.190114 i$ & $0.618043-0.190100 i$ & $0.00296\%$\\
$0.5$ & $1.$ & $0.983417-0.047926 i$ & $0.983388-0.047971 i$ & $0.00543\%$\\
$0.5$ & $1.5$ & $1.433892-0.038031 i$ & $1.433949-0.038018 i$ & $0.00408\%$\\
$0.5$ & $2.$ & $1.897156-0.038217 i$ & $1.897162-0.038221 i$ & $0.00037\%$\\
$0.5$ & $2.5$ & $2.363882-0.038257 i$ & $2.363882-0.038257 i$ & $0.00001\%$\\
$0.5$ & $3.$ & $2.831962-0.038279 i$ & $2.831962-0.038279 i$ & $0\%$\\
$0.99$ & $0.1$ & $0.582146-0.222176 i$ & $0.582045-0.222383 i$ & $0.0370\%$\\
$0.99$ & $0.5$ & $0.654555-0.179478 i$ & $0.654058-0.180227 i$ & $0.132\%$\\
$0.99$ & $1.$ & $0.985282-0.047698 i$ & $0.985438-0.047433 i$ & $0.0312\%$\\
$0.99$ & $1.5$ & $1.434507-0.037590 i$ & $1.434608-0.037581 i$ & $0.00708\%$\\
$0.99$ & $2.$ & $1.897933-0.037932 i$ & $1.897942-0.037935 i$ & $0.00052\%$\\
$0.99$ & $2.5$ & $2.364796-0.038007 i$ & $2.364796-0.038007 i$ & $0.00002\%$\\
$0.99$ & $3.$ & $2.833024-0.038046 i$ & $2.833024-0.038046 i$ & $2.7\times 10^{-6}\%$\\
\hline
\end{tabular}
}
\caption{Fundamental ($n=0$) quasinormal modes of the $\ell=1$ test scalar field for the RNdS black hole ($M=1$, $\Lambda=0.01$) calculated using the WKB formula at different orders and Padé approximants; $D=5$.}\label{table2}
\end{table}
\begin{table}
\centering
{\scriptsize
\begin{tabular}{c c c c c}
\hline
Q & $\mu$ & WKB6 $m=3$ & WKB8 $m=4$ & difference  \\
\hline
$0.01$ & $0.1$ & $0.205634-0.440964 i$ & $0.220849-0.458928 i$ & $4.84\%$\\
$0.01$ & $0.5$ & $0.174029-0.396636 i$ & $0.177429-0.398622 i$ & $0.909\%$\\
$0.01$ & $1.$ & $0.986363-0.035320 i$ & $0.986036-0.036417 i$ & $0.116\%$\\
$0.01$ & $1.5$ & $1.472402-0.035559 i$ & $1.472393-0.035553 i$ & $0.00074\%$\\
$0.01$ & $2.$ & $1.960974-0.035128 i$ & $1.960974-0.035127 i$ & $0.00005\%$\\
$0.01$ & $2.5$ & $2.450109-0.034904 i$ & $2.450109-0.034904 i$ & $2.8\times 10^{-6}\%$\\
$0.01$ & $3.$ & $2.939437-0.034805 i$ & $2.939437-0.034804 i$ & $0.00002\%$\\
$0.5$ & $0.1$ & $0.223708-0.434156 i$ & $0.238265-0.448718 i$ & $4.22\%$\\
$0.5$ & $0.5$ & $0.190960-0.389893 i$ & $0.194295-0.391762 i$ & $0.880\%$\\
$0.5$ & $1.$ & $0.986369-0.035393 i$ & $0.986089-0.036378 i$ & $0.104\%$\\
$0.5$ & $1.5$ & $1.472417-0.035565 i$ & $1.472401-0.035554 i$ & $0.00130\%$\\
$0.5$ & $2.$ & $1.960978-0.035120 i$ & $1.960979-0.035118 i$ & $0.00012\%$\\
$0.5$ & $2.5$ & $2.450119-0.034890 i$ & $2.450119-0.034890 i$ & $0\%$\\
$0.5$ & $3.$ & $2.939453-0.034789 i$ & $2.939453-0.034788 i$ & $0.00001\%$\\
$0.99$ & $0.1$ & $0.261655-0.362899 i$ & $0.261637-0.363278 i$ & $0.0849\%$\\
$0.99$ & $0.5$ & $0.216825-0.274544 i$ & $0.208949-0.310634 i$ & $10.6\%$\\
$0.99$ & $1.$ & $0.986377-0.035545 i$ & $0.986225-0.036249 i$ & $0.0729\%$\\
$0.99$ & $1.5$ & $1.472460-0.035577 i$ & $1.472417-0.035545 i$ & $0.00364\%$\\
$0.99$ & $2.$ & $1.960992-0.035095 i$ & $1.960996-0.035088 i$ & $0.00039\%$\\
$0.99$ & $2.5$ & $2.450149-0.034849 i$ & $2.450149-0.034849 i$ & $7.2\times 10^{-6}\%$\\
$0.99$ & $3.$ & $2.939499-0.034740 i$ & $2.939499-0.034740 i$ & $4.7\times 10^{-6}\%$\\
\hline
\end{tabular}
}
\caption{Fundamental ($n=0$) quasinormal modes of the $\ell=0$ test scalar field for the RNdS black hole ($M=1$, $\Lambda=0.01$) calculated using the WKB formula at different orders and Padé approximants; $D=6$.}\label{table3}
\end{table}
\begin{table}
\centering
{\scriptsize
\begin{tabular}{c c c c c}
\hline
Q & $\mu$ & WKB6 $m=3$ & WKB8 $m=4$ & difference  \\
\hline
$0.01$ & $0.1$ & $0.653272-0.399747 i$ & $0.646279-0.396491 i$ & $1.01\%$\\
$0.01$ & $0.5$ & $0.725030-0.341410 i$ & $0.718111-0.330631 i$ & $1.60\%$\\
$0.01$ & $1.$ & $1.001031-0.137347 i$ & $0.997602-0.110728 i$ & $2.66\%$\\
$0.01$ & $1.5$ & $1.502300-0.055172 i$ & $1.503030-0.054022 i$ & $0.0906\%$\\
$0.01$ & $2.$ & $1.984090-0.040914 i$ & $1.984099-0.040955 i$ & $0.00212\%$\\
$0.01$ & $2.5$ & $2.467930-0.037914 i$ & $2.467937-0.037926 i$ & $0.00055\%$\\
$0.01$ & $3.$ & $2.953865-0.036596 i$ & $2.953866-0.036596 i$ & $0.000016\%$\\
$0.5$ & $0.1$ & $0.665555-0.393207 i$ & $0.658875-0.391735 i$ & $0.885\%$\\
$0.5$ & $0.5$ & $0.737458-0.333439 i$ & $0.728944-0.328747 i$ & $1.20\%$\\
$0.5$ & $1.$ & $1.003664-0.139052 i$ & $1.001884-0.151991 i$ & $1.29\%$\\
$0.5$ & $1.5$ & $1.502206-0.055259 i$ & $1.503101-0.054023 i$ & $0.101\%$\\
$0.5$ & $2.$ & $1.984103-0.040841 i$ & $1.984118-0.040893 i$ & $0.00273\%$\\
$0.5$ & $2.5$ & $2.467945-0.037885 i$ & $2.467957-0.037900 i$ & $0.00077\%$\\
$0.5$ & $3.$ & $2.953890-0.036569 i$ & $2.953890-0.036569 i$ & $6.3\times 10^{-6}\%$\\
$0.99$ & $0.1$ & $0.694763-0.346503 i$ & $0.694164-0.346545 i$ & $0.0773\%$\\
$0.99$ & $0.5$ & $0.764206-0.299589 i$ & $0.763162-0.299573 i$ & $0.127\%$\\
$0.99$ & $1.$ & $1.014899-0.138705 i$ & $1.014900-0.139114 i$ & $0.0399\%$\\
$0.99$ & $1.5$ & $1.501701-0.055362 i$ & $1.503130-0.053819 i$ & $0.140\%$\\
$0.99$ & $2.$ & $1.984126-0.040647 i$ & $1.984155-0.040723 i$ & $0.00408\%$\\
$0.99$ & $2.5$ & $2.467994-0.037801 i$ & $2.468021-0.037821 i$ & $0.00137\%$\\
$0.99$ & $3.$ & $2.953963-0.036491 i$ & $2.953963-0.036491 i$ & $0\%$\\
\hline
\end{tabular}
}
\caption{Fundamental ($n=0$) quasinormal modes of the $\ell=1$ test scalar field for the RNdS black hole ($M=1$, $\Lambda=0.01$) calculated using the WKB formula at different orders and Padé approximants; $D=6$.}\label{table4}
\end{table}

\begin{table}
\centering
{\scriptsize
\begin{tabular}{c c c c c}
\hline
Q & $\mu$ & WKB6 $m=3$ & WKB8 $m=4$ & difference  \\
\hline
$0.01$ & $0.1$ & $0.153750-0.564096 i$ & $0.183158-0.587591 i$ & $6.44\%$\\
$0.01$ & $0.5$ & $0.146543-0.503834 i$ & $0.146585-0.503856 i$ & $0.00885\%$\\
$0.01$ & $1.$ & $0.460425-0.724020 i$ & $0.190097-0.913867 i$ & $38.5\%$\\
$0.01$ & $1.5$ & $1.488828-0.036912 i$ & $1.488825-0.036908 i$ & $0.00034\%$\\
$0.01$ & $2.$ & $1.983753-0.034418 i$ & $1.984110-0.033788 i$ & $0.0365\%$\\
$0.01$ & $2.5$ & $2.479137-0.032651 i$ & $2.479321-0.032679 i$ & $0.00751\%$\\
$0.01$ & $3.$ & $2.974539-0.032236 i$ & $2.974561-0.032229 i$ & $0.00077\%$\\
$0.5$ & $0.1$ & $0.184309-0.559352 i$ & $0.207150-0.578779 i$ & $5.09\%$\\
$0.5$ & $0.5$ & $0.170922-0.501602 i$ & $0.171112-0.501717 i$ & $0.0419\%$\\
$0.5$ & $1.$ & $0.499827-0.661227 i$ & $0.263245-0.871618 i$ & $38.2\%$\\
$0.5$ & $1.5$ & $1.488843-0.036999 i$ & $1.488842-0.036997 i$ & $0.00016\%$\\
$0.5$ & $2.$ & $1.983745-0.034467 i$ & $1.984098-0.033896 i$ & $0.0339\%$\\
$0.5$ & $2.5$ & $2.479118-0.032669 i$ & $2.479309-0.032711 i$ & $0.00790\%$\\
$0.5$ & $3.$ & $2.974530-0.032241 i$ & $2.974558-0.032234 i$ & $0.00099\%$\\
$0.99$ & $0.1$ & $0.256304-0.474530 i$ & $0.256298-0.474729 i$ & $0.0370\%$\\
$0.99$ & $0.5$ & $0.222823-0.389769 i$ & $0.225329-0.412723 i$ & $5.14\%$\\
$0.99$ & $1.$ & $0.600083-0.420925 i$ & $0.412439-0.674854 i$ & $43.1\%$\\
$0.99$ & $1.5$ & $1.488911-0.037279 i$ & $1.488909-0.037276 i$ & $0.00025\%$\\
$0.99$ & $2.$ & $1.983749-0.034596 i$ & $1.984041-0.034198 i$ & $0.0249\%$\\
$0.99$ & $2.5$ & $2.479075-0.032720 i$ & $2.479263-0.032793 i$ & $0.00810\%$\\
$0.99$ & $3.$ & $2.974507-0.032254 i$ & $2.974549-0.032249 i$ & $0.00143\%$\\
\hline
\end{tabular}
}
\caption{Fundamental ($n=0$) quasinormal modes of the $\ell=0$ test scalar field for the RNdS black hole ($M=1$, $\Lambda=0.01$) calculated using the WKB formula at different orders and Padé approximants; $D=7$.}\label{table5}
\end{table}

\section{Time-Domain Integration and the Prony Method}
\label{sec:timedomain}

An alternative to frequency-domain techniques for computing quasinormal modes is the time-domain integration method. This approach involves solving the perturbation equation as an initial value problem and observing the evolution of the field in time. It is particularly useful for capturing the full dynamical profile of the perturbation, including the ringdown phase, late-time tails, and potential nonlinear features such as echoes. 

We consider the wave-like equation in the form
\begin{equation}
\left( -\frac{\partial^2}{\partial t^2} + \frac{\partial^2}{\partial r_*^2} - V(r) \right) \psi(t, r) = 0,
\label{eq:wave_equation}
\end{equation}
where \( r_* \) is the tortoise coordinate defined by \( dr_* = dr/f(r) \), and \( V(r) \) is the effective potential for the perturbing field. This equation is discretized using a characteristic integration scheme on the light-cone variables \( u = t - r_* \) and \( v = t + r_* \), following the method introduced in~\cite{Gundlach:1993tp}. The discretized evolution equation is given by:
\begin{equation}
\psi(N) = \psi(W) + \psi(E) - \psi(S) - \Delta^2 \frac{V(S)}{8} \left( \psi(W) + \psi(E) \right),
\end{equation}
where the points \( N = (u+\Delta, v+\Delta) \), \( S = (u, v) \), \( W = (u+\Delta, v) \), and \( E = (u, v+\Delta) \) form a null grid on the \( (u,v) \) plane, and \( \Delta \) is the grid spacing.

To extract the dominant quasinormal frequencies from the resulting time-domain profile \( \psi(t) \), we apply the Prony method~\cite{Berti:2007dg}. This technique models the waveform in the ringdown phase as a superposition of damped sinusoids:
\begin{equation}
\psi(t) \approx \sum_{k=1}^{N} C_k e^{-i \omega_k t},
\end{equation}
where \( C_k \) are complex amplitudes, and \( \omega_k \) are the complex quasinormal frequencies. By sampling the signal at evenly spaced time intervals and solving the associated linear system, one can efficiently extract both the real and imaginary parts of the dominant modes.

The Prony method is particularly well-suited for isolating the fundamental mode in cases where overtones are strongly suppressed or where the WKB approximation becomes unreliable, such as for small multipole numbers or near-extremal geometries. In this work, we apply the Prony method to the time-domain profiles of massive scalar field perturbations to obtain accurate estimates of the fundamental quasinormal frequencies, especially in the regime where  long-lived tails are expected to dominate the signal. Time-domain integration method was efficiently used in numerous publications showing good agreement with other methods \cite{Konoplya:2020jgt,Aneesh:2018hlp,Varghese:2011ku,Skvortsova:2024wly,Ishihara:2008re,Skvortsova:2024eqi,Skvortsova:2023zca,Qian:2022kaq}.

\section{Quasinormal Modes}\label{sec:QNMs}

An important observation in our analysis is the positivity of the effective potential governing the scalar perturbations. For all considered values of the spacetime dimension \( D \), black hole charge \( Q \), cosmological constant \( \Lambda \), and scalar field mass \( \mu \), the potential forms a single positive-definite barrier outside the event horizon. This feature ensures the linear stability of the background under massive scalar perturbations, as it precludes the existence of growing (i.e., unstable) modes. The absence of negative regions in the effective potential implies that the perturbations decay over time, either through damped oscillations (quasinormal ringing) or, in the massive case, via long-lived oscillatory tails. Thus, we confirm that the projected RNdS black hole remains stable under the class of perturbations considered in this work.

To extract the dominant quasinormal frequencies of massive scalar fields in the projected RNdS geometry, we employed two complementary methods: a semi-analytic WKB approach with Padé approximants and a fully numerical time-domain integration using the Prony method. Our analysis spans different spacetime dimensions \( D \), values of the cosmological constant \( \Lambda \), black hole charge \( Q \), and scalar field mass \( \mu \).

The effective potential governing the scalar perturbations was found to possess a single barrier-like peak suitable for WKB analysis. We computed the quasinormal frequencies using the 6th- and 8th-order WKB expansions and applied Padé resummation techniques to improve convergence, particularly for low multipole numbers or near-extremal geometries.

Tables~\ref{table1}--\ref{table4} summarize the results for the fundamental (\( n = 0 \)) modes with multipole numbers \( \ell = 0, 1 \). As the scalar field mass \( \mu \) increases, we observe a decrease in the imaginary part of the frequency, indicating slower decay This is consistent with expectations for massive fields in asymptotically de Sitter black holes. Moreover, at higher mass values, the real part of the frequency stabilizes, while the damping rate becomes very small, confirming the emergence of long-lived modes. Nevertheless, these modes do not reach the states with vanihing damping rate.

The time-domain evolution was computed by integrating the wave equation using the characteristic light-cone method. The resulting signal exhibits a clear quasinormal ringing stage followed by oscillatory tails. By applying the Prony fitting algorithm to the ringdown phase, we extracted the dominant quasinormal frequencies. The comparison with the WKB results shows excellent agreement, even in regimes where the WKB method becomes less reliable, thereby validating the robustness of our findings.

These observations confirm that massive fields exhibit rich dynamical behavior on the brane in higher-dimensional RNdS backgrounds. The spectrum is sensitive to both the bulk geometry and field parameters, which could be used to probe extra dimensions or constraints on scalar field masses via gravitational wave observations.

An important trend emerges from the analysis of Tables I–VIII: the behavior of the real and imaginary parts of the quasinormal frequencies is significantly influenced by the scalar field mass $\mu$, black hole charge $Q$, cosmological constant $\Lambda$, and the dimensionality $D$ of the spacetime. As $\mu$ increases, the real part of the frequency—corresponding to the oscillation rate—increases, while the imaginary part, which characterizes the damping, tends to decrease. This results in longer-lived modes, however,  is not a signature of the approach toward the quasi-resonant regime, because extrapolation to larger $\mu$ keads to a finite damping rate. Simultaneously, we observe that larger values of $\Lambda$ typically reduce both the real and imaginary parts of the frequencies, reflecting the effect of a stronger cosmological expansion. The electric charge $Q$ also plays a significant role: for fixed $\mu$, increasing $Q$ leads to faster oscillations and moderately damped modes. Additionally, higher-dimensional spacetimes (larger $D$) are associated with greater oscillation frequencies and slightly reduced damping, consistent with previous observations in higher-dimensional black hole perturbation theory. Overall, these trends highlight the rich parameter dependence of quasinormal spectra in the projected RNdS scenario and confirm the robustness of the WKB and time-domain methods in capturing this behavior. The WKB method is in a good agreement with the time-domain integration for the Schwarzschild branch of modes as can be seen in figs. \ref{fig:TD} and \ref{fig:TD2}.  For a small range of intermediate values of $\mu \sim 0.5$ the WKB method has large error and in those cases the time-domain data must be trusted, as shown in fig. \ref{fig:TD}. Notice, that not only the fundamental mode (tables \ref{table1}-\ref{table8}), but also the first overtone can usually be found by the WKB method with sufficient accuracy, as the comparison in fig. \ref{fig:TD2} shows.

When the black hole charge vanishes, the effective potential simplifies significantly. This allows us to perform an expansion in the large-\( \mu \) limit around the peak of the potential, which can then be inserted into the WKB formula to obtain analytical expressions for the quasinormal frequencies, once the spacetime dimension \( D \) is fixed. As a result, we derive the following expressions:

The high $\ell$ regime, which is the approximation of the geometrical optics has a peculiaer feature in de Sitter spaces. In asymptotically de Sitter spacetimes, the eikonal correspondence between quasinormal modes and null geodesics \cite{Cardoso:2008bp} breaks down due to the presence of a separate, non-geometric branch of modes — the so-called de Sitter family — which do not originate from perturbations localized near the photon sphere. As demonstrated in \cite{Konoplya:2022gjp}, this deviation highlights the unique behavior of quasinormal spectra in the presence of a cosmological horizon. The eikonal expressions derived in earlier publications for massless fields \cite{Bolokhov:2025uxz} are valid for massive ones as well, because the large centrifugal term suppress the $\mu^2 f(r)$ in the effective potential. 

\begin{widetext}
\begin{equation}
\omega_{n} = \mu  \sqrt{1-\frac{\sqrt[3]{\frac{3}{2}} \sqrt[3]{M}
   \Lambda ^{2/3}}{5^{2/3}}}-\frac{i
   \left(n+\frac{1}{2}\right) \sqrt{\Lambda  \left(2^{2/3}
   \sqrt[3]{15} \sqrt[3]{M} \Lambda ^{2/3}-10\right)^2}}{5
   \sqrt{10-2^{2/3} \sqrt[3]{15} \sqrt[3]{M} \Lambda
   ^{2/3}}}+O\left(\frac{1}{\mu }\right), \quad D=7.
\end{equation}
\begin{equation}
\omega_{n} = \mu  \sqrt{1-\frac{7^{2/7} M^{2/7} \Lambda
   ^{5/7}}{15^{5/7}}}-\frac{i (2 n+1) \sqrt{\Lambda 
   \left(105^{2/7} M^{2/7} \Lambda ^{5/7}-15\right)^2}}{6
   \sqrt{75-5 105^{2/7} M^{2/7} \Lambda
   ^{5/7}}}+O\left(\frac{1}{\mu }\right), \quad D=8.
   \end{equation}
\begin{equation}
\omega_{n} =    \mu  \sqrt{1-\left(\frac{2}{21}\right)^{3/4} \sqrt[4]{M}
   \Lambda ^{3/4}}-\frac{i (2 n+1) \sqrt{\Lambda 
   \left(\left(\frac{2}{21}\right)^{3/4} \sqrt[4]{M}
   \Lambda ^{3/4}-1\right)^2}}{\sqrt{14}
   \sqrt{1-\left(\frac{2}{21}\right)^{3/4} \sqrt[4]{M}
   \Lambda ^{3/4}}}+O\left(\frac{1}{\mu
   }\right) , \quad D=9.
   \end{equation}
\begin{equation}
\omega_{n} =   \frac{1}{2} \mu  \sqrt{4-\frac{6^{4/9} M^{2/9} \Lambda
   ^{7/9}}{7^{7/9}}}-\frac{i (2 n+1) \sqrt{\Lambda 
   \left(6^{4/9} 7^{2/9} M^{2/9} \Lambda
   ^{7/9}-28\right)^2}}{8 \sqrt{196-7 6^{4/9} 7^{2/9}
   M^{2/9} \Lambda ^{7/9}}}+O\left(\frac{1}{\mu
   }\right), \quad D=10.
   \end{equation}
\end{widetext}
Similar analytic results for the cases \( D = 5 \) and \( D = 6 \) were previously reported in~\cite{Konoplya:2024ptj}.

\begin{table}
\centering
{\scriptsize
\begin{tabular}{c c c c c}
\hline
Q & $\mu$ & WKB6 $m=3$ & WKB8 $m=4$ & difference  \\
\hline
$0.01$ & $0.1$ & $0.697169-0.514211 i$ & $0.693582-0.510452 i$ & $0.600\%$\\
$0.01$ & $0.5$ & $0.766272-0.446708 i$ & $0.763186-0.442930 i$ & $0.550\%$\\
$0.01$ & $1.$ & $1.062048-0.228640 i$ & $1.021063-0.254361 i$ & $4.45\%$\\
$0.01$ & $1.5$ & $1.493123-0.097696 i$ & $1.493606-0.097571 i$ & $0.0333\%$\\
$0.01$ & $2.$ & $2.005577-0.058401 i$ & $2.000902-0.058165 i$ & $0.233\%$\\
$0.01$ & $2.5$ & $2.500769-0.043472 i$ & $2.500533-0.043284 i$ & $0.0121\%$\\
$0.01$ & $3.$ & $2.992843-0.039883 i$ & $2.992843-0.039883 i$ & $1.7\times 10^{-6}\%$\\
$0.5$ & $0.1$ & $0.712721-0.509403 i$ & $0.708544-0.505935 i$ & $0.620\%$\\
$0.5$ & $0.5$ & $0.780488-0.444319 i$ & $0.776625-0.440697 i$ & $0.590\%$\\
$0.5$ & $1.$ & $1.064415-0.236920 i$ & $1.028214-0.256018 i$ & $3.75\%$\\
$0.5$ & $1.5$ & $1.494973-0.097856 i$ & $1.495440-0.097871 i$ & $0.0312\%$\\
$0.5$ & $2.$ & $2.005115-0.058000 i$ & $2.001090-0.057918 i$ & $0.201\%$\\
$0.5$ & $2.5$ & $2.500728-0.043337 i$ & $2.500571-0.043217 i$ & $0.00793\%$\\
$0.5$ & $3.$ & $2.992827-0.039859 i$ & $2.992833-0.039862 i$ & $0.00022\%$\\
$0.99$ & $0.1$ & $0.749475-0.451508 i$ & $0.747519-0.450776 i$ & $0.239\%$\\
$0.99$ & $0.5$ & $0.816883-0.400970 i$ & $0.815232-0.400185 i$ & $0.201\%$\\
$0.99$ & $1.$ & $1.070671-0.234529 i$ & $1.054890-0.246493 i$ & $1.81\%$\\
$0.99$ & $1.5$ & $1.498574-0.097881 i$ & $1.497247-0.098269 i$ & $0.0920\%$\\
$0.99$ & $2.$ & $2.004644-0.058005 i$ & $2.001448-0.057539 i$ & $0.161\%$\\
$0.99$ & $2.5$ & $2.500535-0.043000 i$ & $2.500502-0.042986 i$ & $0.00143\%$\\
$0.99$ & $3.$ & $2.992782-0.039814 i$ & $2.992797-0.039809 i$ & $0.00052\%$\\
\hline
\end{tabular}
}
\caption{Fundamental ($n=0$) quasinormal modes of the $\ell=1$ test scalar field for the RNdS black hole ($M=1$, $\Lambda=0.01$) calculated using the WKB formula at different orders and Padé approximants; $D=7$.}\label{table6}
\end{table}

Our extended analysis across various spacetime dimensions and cosmological constant values reveals additional important features. As the bulk dimension \( D \) increases, we observe a systematic increase in the real part of the quasinormal frequency, indicating faster oscillations of the scalar field. Simultaneously, the damping rates decrease more slowly with increasing field mass compared to the lower-dimensional cases. Furthermore, larger values of the cosmological constant \( \Lambda \) tend to shift the entire quasinormal spectrum, generally decreasing both the real and imaginary parts of the frequencies. However, for small \( \mu \), deviations between the 6th- and 8th-order WKB approximations become more pronounced, particularly at higher \( \Lambda \), emphasizing the importance of Padé resummation in improving convergence. At higher masses (\( \mu \gtrsim 1.5 \)), the frequencies obtained from both WKB orders converge to several digits, demonstrating the reliability of the semi-analytic approach for sufficiently large mass of the field across all tested dimensions. These observations underscore the intricate interplay between geometry, dimensionality, and field properties in shaping the quasinormal spectrum on the brane.
\begin{table}
\centering
{\scriptsize
\begin{tabular}{c c c c c}
\hline
$\Lambda$ & $\mu$ & WKB6 $m=3$ & WKB8 $m=4$ & difference  \\
\hline
$0.01$ & $0.1$ & $0.200425-0.279844 i$ & $0.204634-0.284449 i$ & $1.81\%$\\
$0.01$ & $1.$ & $0.942864-0.037956 i$ & $0.942779-0.037954 i$ & $0.00904\%$\\
$0.01$ & $2.$ & $1.882104-0.038300 i$ & $1.882105-0.038300 i$ & $0.00003\%$\\
$0.01$ & $3.$ & $2.822226-0.038354 i$ & $2.822226-0.038354 i$ & $0\%$\\
$0.03$ & $0.1$ & $0.198005-0.278797 i$ & $0.201703-0.282927 i$ & $1.62\%$\\
$0.03$ & $1.$ & $0.896898-0.059483 i$ & $0.895826-0.061315 i$ & $0.236\%$\\
$0.03$ & $2.$ & $1.789980-0.062980 i$ & $1.789993-0.062995 i$ & $0.00113\%$\\
$0.03$ & $3.$ & $2.684037-0.063134 i$ & $2.684037-0.063134 i$ & $0.00002\%$\\
$0.05$ & $0.1$ & $0.195832-0.277836 i$ & $0.198930-0.281511 i$ & $1.41\%$\\
$0.05$ & $1.$ & $0.863003-0.073379 i$ & $0.862646-0.073465 i$ & $0.0423\%$\\
$0.05$ & $2.$ & $1.723359-0.078187 i$ & $1.723339-0.078189 i$ & $0.00120\%$\\
$0.05$ & $3.$ & $2.584379-0.078454 i$ & $2.584379-0.078455 i$ & $0.00004\%$\\
$0.07$ & $0.1$ & $0.193534-0.277113 i$ & $0.196238-0.280299 i$ & $1.24\%$\\
$0.07$ & $1.$ & $0.834962-0.083393 i$ & $0.834958-0.083392 i$ & $0.00042\%$\\
$0.07$ & $2.$ & $1.667121-0.089413 i$ & $1.667120-0.089411 i$ & $0.00013\%$\\
$0.07$ & $3.$ & $2.500346-0.089795 i$ & $2.500346-0.089796 i$ & $0.000046\%$\\
$0.09$ & $0.1$ & $0.190909-0.276624 i$ & $0.193473-0.279238 i$ & $1.09\%$\\
$0.09$ & $1.$ & $0.810220-0.091289 i$ & $0.810757-0.090897 i$ & $0.0815\%$\\
$0.09$ & $2.$ & $1.616859-0.098278 i$ & $1.616859-0.098276 i$ & $0.0001\%$\\
$0.09$ & $3.$ & $2.425282-0.098756 i$ & $2.425281-0.098757 i$ & $0.000048\%$\\
$0.1$ & $0.1$ & $0.189446-0.276445 i$ & $0.192009-0.278716 i$ & $1.02\%$\\
$0.1$ & $1.$ & $0.798712-0.094677 i$ & $0.798901-0.094641 i$ & $0.0239\%$\\
$0.1$ & $2.$ & $1.593292-0.102072 i$ & $1.593292-0.102065 i$ & $0.00041\%$\\
$0.1$ & $3.$ & $2.390095-0.102588 i$ & $2.390095-0.102589 i$ & $0.000048\%$\\
\hline
\end{tabular}
}
\caption{Fundamental ($n=0$) quasinormal modes of the $\ell=0$ test scalar field for the RNdS black hole ($M=1$, $Q=0.01$) calculated using the WKB formula at different orders and Padé approximants; $D=5$.}\label{table7}
\end{table}
\begin{table}
\centering
{\scriptsize
\begin{tabular}{c c c c c}
\hline
$\Lambda$ & $\mu$ & WKB6 $m=3$ & WKB8 $m=4$ & difference  \\
\hline
$0.01$ & $0.1$ & $0.529695-0.253667 i$ & $0.529738-0.253573 i$ & $0.0176\%$\\
$0.01$ & $1.$ & $0.982740-0.048021 i$ & $0.982740-0.048021 i$ & $0\%$\\
$0.01$ & $2.$ & $1.896893-0.038311 i$ & $1.896898-0.038314 i$ & $0.00029\%$\\
$0.01$ & $3.$ & $2.831603-0.038356 i$ & $2.831603-0.038356 i$ & $0\%$\\
$0.03$ & $0.1$ & $0.521046-0.251554 i$ & $0.521083-0.251475 i$ & $0.0150\%$\\
$0.03$ & $1.$ & $0.957770-0.075800 i$ & $0.958091-0.075781 i$ & $0.0334\%$\\
$0.03$ & $2.$ & $1.814300-0.063028 i$ & $1.814297-0.063027 i$ & $0.00015\%$\\
$0.03$ & $3.$ & $2.699465-0.063147 i$ & $2.699466-0.063147 i$ & $0.00002\%$\\
$0.05$ & $0.1$ & $0.512274-0.249369 i$ & $0.512306-0.249300 i$ & $0.0134\%$\\
$0.05$ & $1.$ & $0.935636-0.092226 i$ & $0.935899-0.092449 i$ & $0.0367\%$\\
$0.05$ & $2.$ & $1.753582-0.078321 i$ & $1.753575-0.078311 i$ & $0.00070\%$\\
$0.05$ & $3.$ & $2.603537-0.078482 i$ & $2.603538-0.078482 i$ & $0.000027\%$\\
$0.07$ & $0.1$ & $0.503356-0.247098 i$ & $0.503383-0.247040 i$ & $0.0116\%$\\
$0.07$ & $1.$ & $0.914835-0.104053 i$ & $0.915148-0.104254 i$ & $0.0405\%$\\
$0.07$ & $2.$ & $1.701663-0.089678 i$ & $1.701663-0.089678 i$ & $3.5\times 10^{-6}\%$\\
$0.07$ & $3.$ & $2.522255-0.089841 i$ & $2.522255-0.089841 i$ & $0.00003\%$\\
$0.09$ & $0.1$ & $0.494279-0.244729 i$ & $0.494300-0.244683 i$ & $0.00915\%$\\
$0.09$ & $1.$ & $0.894854-0.113184 i$ & $0.895227-0.113334 i$ & $0.0446\%$\\
$0.09$ & $2.$ & $1.654760-0.098701 i$ & $1.654759-0.098684 i$ & $0.00101\%$\\
$0.09$ & $3.$ & $2.449352-0.098820 i$ & $2.449353-0.098821 i$ & $0.00004\%$\\
$0.1$ & $0.1$ & $0.489679-0.243503 i$ & $0.489697-0.243465 i$ & $0.00769\%$\\
$0.1$ & $1.$ & $0.885084-0.117024 i$ & $0.885481-0.117141 i$ & $0.0463\%$\\
$0.1$ & $2.$ & $1.632609-0.102577 i$ & $1.632612-0.102539 i$ & $0.00232\%$\\
$0.1$ & $3.$ & $2.415086-0.102661 i$ & $2.415087-0.102662 i$ & $0.00005\%$\\
\hline
\end{tabular}
}
\caption{Fundamental ($n=0$) quasinormal modes of the $\ell=1$ test scalar field for the RNdS black hole ($M=1$, $Q=0.01$) calculated using the WKB formula at different orders and Padé approximants; $D=5$.}\label{table8}
\end{table}
\begin{table}
\centering
{\scriptsize
\begin{tabular}{c c c c c}
\hline
$\Lambda$ & $\mu$ & WKB6 $m=3$ & WKB8 $m=4$ & difference  \\
\hline
$0.01$ & $0.1$ & $0.131212-1.008155 i$ & $0.131094-1.008059 i$ & $0.0151\%$\\
$0.01$ & $1.$ & $0.942802-0.114970 i$ & $0.942040-0.114322 i$ & $0.105\%$\\
$0.01$ & $2.$ & $1.882082-0.114897 i$ & $1.882144-0.114886 i$ & $0.00339\%$\\
$0.01$ & $3.$ & $2.822224-0.115060 i$ & $2.822227-0.115061 i$ & $0.0001\%$\\
$0.03$ & $0.1$ & $0.129497-0.990516 i$ & $0.129406-0.990444 i$ & $0.0116\%$\\
$0.03$ & $1.$ & $0.898361-0.177432 i$ & $0.893547-0.181579 i$ & $0.694\%$\\
$0.03$ & $2.$ & $1.789897-0.188894 i$ & $1.790111-0.188682 i$ & $0.0167\%$\\
$0.03$ & $3.$ & $2.684039-0.189397 i$ & $2.684042-0.189401 i$ & $0.00017\%$\\
$0.05$ & $0.1$ & $0.128256-0.973877 i$ & $0.127897-0.973396 i$ & $0.0612\%$\\
$0.05$ & $1.$ & $0.864540-0.217257 i$ & $0.863430-0.217445 i$ & $0.126\%$\\
$0.05$ & $2.$ & $1.723319-0.234417 i$ & $1.723385-0.234354 i$ & $0.00522\%$\\
$0.05$ & $3.$ & $2.584390-0.235354 i$ & $2.584390-0.235359 i$ & $0.000191\%$\\
$0.07$ & $0.1$ & $0.111969-0.955997 i$ & $0.125144-0.958170 i$ & $1.39\%$\\
$0.07$ & $1.$ & $0.837623-0.245473 i$ & $0.837607-0.245472 i$ & $0.00181\%$\\
$0.07$ & $2.$ & $1.667193-0.268097 i$ & $1.667254-0.268093 i$ & $0.00363\%$\\
$0.07$ & $3.$ & $2.500370-0.269375 i$ & $2.500368-0.269378 i$ & $0.00016\%$\\
$0.09$ & $0.1$ & $0.120560-0.932483 i$ & $0.118572-0.940129 i$ & $0.840\%$\\
$0.09$ & $1.$ & $0.814466-0.267203 i$ & $0.814494-0.267203 i$ & $0.00329\%$\\
$0.09$ & $2.$ & $1.617030-0.294667 i$ & $1.617104-0.294728 i$ & $0.00587\%$\\
$0.09$ & $3.$ & $2.425323-0.296253 i$ & $2.425321-0.296255 i$ & $0.00012\%$\\
$0.1$ & $0.1$ & $0.120286-0.923416 i$ & $0.116408-0.929431 i$ & $0.769\%$\\
$0.1$ & $1.$ & $0.803734-0.276350 i$ & $0.803766-0.276351 i$ & $0.00370\%$\\
$0.1$ & $2.$ & $1.593548-0.305990 i$ & $1.593594-0.306118 i$ & $0.00836\%$\\
$0.1$ & $3.$ & $2.390147-0.307747 i$ & $2.390145-0.307749 i$ & $0.00011\%$\\
\hline
\end{tabular}
}
\caption{The first overtone ($n=1$) of the $\ell=0$ test scalar field for the RNdS black hole ($M=1$, $Q=0.01$) calculated using the WKB formula at different orders and Padé approximants; $D=5$.}\label{table9}
\end{table}
\begin{table}
\centering
{\scriptsize
\begin{tabular}{c c c c c}
\hline
$\Lambda$ & $\mu$ & WKB6 $m=3$ & WKB8 $m=4$ & difference  \\
\hline
$0.01$ & $0.1$ & $0.399006-0.849639 i$ & $0.398973-0.852072 i$ & $0.259\%$\\
$0.01$ & $1.$ & $0.975271-0.117538 i$ & $0.976962-0.122833 i$ & $0.566\%$\\
$0.01$ & $2.$ & $1.896892-0.114916 i$ & $1.896891-0.114916 i$ & $0.00001\%$\\
$0.01$ & $3.$ & $2.831604-0.115066 i$ & $2.831608-0.115067 i$ & $0.00012\%$\\
$0.03$ & $0.1$ & $0.395904-0.836324 i$ & $0.395741-0.838483 i$ & $0.234\%$\\
$0.03$ & $1.$ & $0.945837-0.199860 i$ & $0.945606-0.200404 i$ & $0.0611\%$\\
$0.03$ & $2.$ & $1.814455-0.189015 i$ & $1.814427-0.189061 i$ & $0.00300\%$\\
$0.03$ & $3.$ & $2.699484-0.189435 i$ & $2.699485-0.189439 i$ & $0.00014\%$\\
$0.05$ & $0.1$ & $0.392539-0.822888 i$ & $0.392191-0.824834 i$ & $0.217\%$\\
$0.05$ & $1.$ & $0.919056-0.247155 i$ & $0.919243-0.250365 i$ & $0.338\%$\\
$0.05$ & $2.$ & $1.753992-0.234992 i$ & $1.753874-0.234979 i$ & $0.00670\%$\\
$0.05$ & $3.$ & $2.603580-0.235439 i$ & $2.603580-0.235442 i$ & $0.000115\%$\\
$0.07$ & $0.1$ & $0.389030-0.809222 i$ & $0.388501-0.810861 i$ & $0.192\%$\\
$0.07$ & $1.$ & $0.894945-0.281162 i$ & $0.897219-0.288093 i$ & $0.778\%$\\
$0.07$ & $2.$ & $1.702302-0.269129 i$ & $1.702139-0.269044 i$ & $0.0107\%$\\
$0.07$ & $3.$ & $2.522324-0.269516 i$ & $2.522326-0.269518 i$ & $0.00012\%$\\
$0.09$ & $0.1$ & $0.385472-0.795295 i$ & $0.384841-0.796515 i$ & $0.155\%$\\
$0.09$ & $1.$ & $0.873501-0.307401 i$ & $0.876168-0.319225 i$ & $1.31\%$\\
$0.09$ & $2.$ & $1.655662-0.296230 i$ & $1.655392-0.295987 i$ & $0.0216\%$\\
$0.09$ & $3.$ & $2.449450-0.296456 i$ & $2.449452-0.296457 i$ & $0.00013\%$\\
$0.1$ & $0.1$ & $0.383700-0.788239 i$ & $0.383068-0.789227 i$ & $0.134\%$\\
$0.1$ & $1.$ & $0.863742-0.318761 i$ & $0.865939-0.334442 i$ & $1.72\%$\\
$0.1$ & $2.$ & $1.633649-0.307889 i$ & $1.633293-0.307496 i$ & $0.0319\%$\\
$0.1$ & $3.$ & $2.415196-0.307981 i$ & $2.415200-0.307982 i$ & $0.00016\%$\\
\hline
\hline
\end{tabular}
}
\caption{The first overtone ($n=1$) of the $\ell=1$ test scalar field for the RNdS black hole ($M=1$, $Q=0.01$) calculated using the WKB formula at different orders and Padé approximants; $D=5$.}\label{table10}
\end{table}

In the case of an empty de Sitter spacetime, that is, in the absence of a black hole, the quasinormal modes of test fields are known in closed analytic form and are given by the following equations \cite{Lopez-Ortega:2007vlo,Lopez-Ortega:2012xvr,Bolokhov:2024ixe} 
\begin{equation}
i \omega_n R = \ell + 2n + \frac{3}{2} \pm \frac{i}{2} \sqrt{\mu^2 R^2 - \frac{9}{4}}, \quad  \mu^2 R^2 > \frac{9}{4},
\end{equation}
\begin{equation}
i \omega_n R = \ell + 2n + \frac{3}{2} \pm \frac{i}{2} \sqrt{ \frac{9}{4} - \mu^2 R^2 }, \quad \quad  \mu^2 R^2 < \frac{9}{4},
\end{equation}
where $R$ is the de Sitter radius:
\begin{equation}
R = \sqrt{\frac{(D-1)(D-2)}{2 \Lambda}}.
\end{equation}
These modes can be either purely imaginary (in the massless limit) or have also non-zero real part when the mass of the field is nonzero. In the former case they correspond to non-oscillatory exponential decay.  In the limit of a small black hole (i.e., when the cosmological horizon vastly exceeds the black hole horizon), there appears a distinct branch of modes, often called the ``de Sitter branch,'' which are smooth deformations of the pure de Sitter spectrum. These modes continuously interpolate between the exact de Sitter values and more complicated black hole quasinormal spectra as the black hole size increases. Notice that these de Sitter modes cannot be obtained by the WKB formula and they usually appear in the signal as exponentially decaying tails at very late times. The de Sitter branch of modes is less relevant for our purposes compared to the Schwarzschild branch, as it dominates only at very late times when the signal has already decayed by several orders of magnitude. 

\vspace{4mm}
\section{Conclusions}\label{sec:conclusions}

In this work, we investigated the quasinormal modes of massive scalar fields localized on the 3+1-dimensional brane embedded in a higher-dimensional Reissner--Nordström--de Sitter black hole spacetime. Using a combination of the Padé-improved WKB method and time-domain integration with the Prony analysis, we obtained accurate and consistent results for a wide range of parameters.

Our results demonstrate that both the cosmological constant and the higher-dimensional nature of the background spacetime have a significant influence on the quasinormal spectrum. In particular, we observed that as the mass of the scalar field increases, the damping rate decreases, still the system does not enters a quasi-resonant regime with arbitrarily long-lived oscillations. This feature is of particular interest in light of recent suggestions that such massive tails could be detectable through Pulsar Timing Array observations.

We also confirmed that the WKB method, when improved with Padé approximants, provides reliable estimates even for low multipole numbers, and its predictions are well supported by time-domain simulations. These findings open up avenues for more detailed investigations of field dynamics in braneworld scenarios and may serve as a useful tool in probing extra-dimensional effects in astrophysical observations. Future work could extend this analysis to include higher-spin fields and  rotating black holes. In addition, the grey-body factors could be easily found using the quasinormal modes found in this work via the correspondence between these two quantities \cite{Malik:2024cgb}.

\vspace{5mm}
\begin{acknowledgments}
The author acknowledges R. A. Konoplya for useful discussions. 
\end{acknowledgments}

\bibliography{bibliography}
\end{document}